\documentclass[seceq]{ptptex}
\usepackage{wrapfig}

\usepackage{amsthm}
\usepackage[dvips]{graphicx}

\newcommand{\bm}[1]{\mbox{\boldmath{$#1$}}}  
\newcommand{\be}{\begin{equation}}
\newcommand{\ee}{\end{equation}}
\newcommand{\ba}{\begin{eqnarray}}
\newcommand{\ea}{\end{eqnarray}}

\newcommand{\nn}{\nonumber}

\notypesetlogo                       

\title{
Effect of the Zero-Mode on the Response of a Trapped Bose-Condensed Gas
}

\author{
Makoto \textsc{Mine},$^{1,}$\footnote{E-mail: mine@toki.waseda.jp
} 
Tomoi \textsc{Koide},$^{2,}$\footnote{E-mail: koide@fma.if.usp.br
}
Masahiko \textsc{Okumura}$^{3,}$\footnote{E-mail: okumura@aoni.waseda.jp
} \\
and Yoshiya \textsc{Yamanaka}$^{4,}$\footnote{E-mail: yamanaka@waseda.jp}
}

\inst{
$^1$Department of Physics, Waseda University,
 Tokyo 169-8555, Japan
\\
$^2$Instituto de F\'{\i}sica, Universidade de S\~ao Paulo, C.P. 66318,
05315-970 S\~ao Paulo-SP, Brazil 
\\
$^3$Department of Applied Physics, Waseda University,
 Tokyo 169-8555, Japan
\\
$^4$Department of Materials Science and Engineering,
Waseda University, Tokyo 169-8555, Japan
}

\recdate{October 26, 2005}

\abst{
The dynamical response of a trapped Bose-Einstein condensate (BEC) is
formulated consistently with quantum field theory and is 
numerically evaluated. 
We regard the BEC as a manifestation of the breaking of the global
phase symmetry.  
Then, the Goldstone theorem implies the existence of a zero energy
excitation mode (the zero-mode).
We calculate the effect of the zero-mode on the response frequency and
show that the contribution of the zero-mode to the first excitation mode 
is not so important in the parameter set realized in the existing experiment.
This is the reason that experimental results can be described 
using the Bogoliubov prescription, although it breaks the 
consistency of the description in quantum field theory. 
} 

\begin{document}

\maketitle

\section{\label{intro}Introduction}

Since the experimental realization of the Bose-Einstein condensate (BEC)
in trapped dilute atomic gases, a large number of works associated with
trapped BECs have been reported. 
The majority of these works are devoted to the analysis of the
Gross-Pitaevskii (GP) equation \cite{GP}.
The GP equation has been successful in accounting for
experimental results at very low temperature\cite{Dalfovo}. 
However, the GP equation is not capable of describing the BECs as the
temperature is raised, because, for instance, we should take the effect
of the non-condensate fractions into account.
Moreover, we can continue to fine-tune experiments
on trapped BECs, 
changing the scattering length between atoms, by utilizing
Feshbach resonance\cite{Feshbach}. 
Thus, trapped BECs afford an opportunity for testing quantum
field theory (QFT), which is still incomplete. 
To this end, we need to analyze trapped BECs beyond the GP equation.

There are several theoretical works in which the 
behavior of BECs at finite temperature both
with\cite{Gri1,Gri2,Hut,Rus,Choi,Kru,JZ,MRHB,Morgan1, Morgan2} 
and without a trapping
potential\cite{Ramos1,Ramos2,Ramos3,Boya} is 
studied. 
In particular, 
the results given in
Ref.~\citen{JZ}, as well as Refs.~\citen{MRHB,Morgan1, Morgan2}, 
are in good agreement with the experimental value\cite{Jin} of 
the excitation frequencies
from the low temperature region to the region near the transition temperature. 
The authors of Ref.~\citen{JZ} use Zaremba--Nikuni--Griffin 
formalism\cite{ZNK}, 
in which the time-dependent GP equation couples with the semiclassical Boltzmann 
equation. 
In Refs.~\citen{MRHB,Morgan1, Morgan2}, 
the authors apply a time-dependent extension of the perturbation theory.

The quantization scheme of a trapped BEC is, however, an open question.
From the viewpoint of QFT, the BEC is a manifestation of the spontaneous
breakdown of a global phase symmetry. 
In this picture, the Goldstone theorem requires the existence of a
mode with zero excitation energy (the zero-mode) for consistency with the
fundamental concepts of QFT. 
However, a consistent quantization scheme is not easy to construct,
because of, for instance, the problem of the infrared divergence. 
Usually, when the Bogoliubov prescription is employed, 
the zero-mode is ingored, but the naive application of the Bogoliubov
prescription gives rise to inconsistency, because the precise
canonical commutation relations cannot be satisfied.
Thus, we cannot ignore the zero-mode in testing the predictions of quantum
field theory in trapped BEC experiments.

Recently, a consistent quantization scheme was proposed by two of the
present authors\cite{OY}. 
They introduced an infinitesimal breaking term of the global phase
symmetry.
Here, the breaking term plays roles as a regulator of infrared
divergences and enables us to construct the generalized Bogoliubov
transformation (GBT) including the zero-mode\cite{OY}. 
It has been proven that this mode corresponds to the Nambu-Goldstone
(NG) mode by demonstrating that the Ward-Takahashi (WT)
relations\cite{OY,OY05} are satisfied. 
Realization of unitarily inequivalent vacua that belong to order
parameters with different phases was shown in Ref.~\citen{OYVac}. 
The existence of such vacua implies that the mechanism of the 
the spontaneous symmetry breakdown works even in
traps. 
There is another consistent field theoretical approach, in which the
operators corresponding to the zero-mode are given by ``quantum
coordinates''\cite{Lewen, Matsu}. 
The relation between the zero-mode approach and the quantum
coordinate approach has been revealed\cite{MOY}, 
but it has not been confirmed that the WT relations are satisfied 
in the latter approach. 
It should also be mentioned that 
no definite criterion has been found for choosing quantum states 
acted on by quantum coordinates.
Effects of the zero-mode, as well as modes represented by the
quantum coordinates on the condensate density have been
evaluated\cite{OY,OY04},  
and, unfortunately, they were found to be too small to be detected 
in present-day experiments.

The aim of the consistent canonical studies cited above 
is to elucidate the equilibrium properties of trapped BECs 
\cite{OY, OY05, OYVac,Lewen,Matsu,MOY,OY04}.
In this paper, we apply the consistent quantization scheme and 
study the effect of the 
zero-mode on the dynamical response 
of a trapped BEC at finite temperature.

This paper is organized as follows. In \S 2, we explain the
quantization of a trapped BEC including the zero-mode. 
In \S 3, the dynamical response of the BEC under an external field is
derived, and the results of numerical calculations are reported
in \S 4. Section~5 contains
concluding remarks. 
We give a brief summary of the mathematics regarding the GBT including
the zero-mode in Appendix A.

\section{Quantization scheme consistent with the Goldstone theorem}\label{sec-Model}

In this section, we give a brief review of the quantization of a
static trapped BEC including the zero-mode, following
Refs. \citen{MOY,OY05} and \citen{OY04}.

We adopt the following action, which describes the dynamics of a neutral gas
system trapped in a potential $V$:
\ba
S[\psi,\psi^{\dagger}] = \int dt\,d^{3}\bm{x}\,
\left[
\psi^{\dag} (x)
\left(
T -K -V + \mu
\right)
\psi(x) 
-\frac{g}{2} \psi^{\dag 2}(x) \psi^2(x)
\right]
 \, . \label{S}
\ea
Here,
$ T = i \frac{\partial}{\partial t}$, 
$ K = - \frac{1}{2m} \nabla^2 $, and $\mu$ and $g$ denote the chemical
potential and the coupling constant, respectively.
We choose the harmonic trapping potential as $ V = \frac{m}{2}
\sum_{i=1}^3 \omega_i^2 x_i^2$, 
where $\omega_i$ is the trapping frequency along the $x_i$ direction. 
The symbol $x$ represents $({\bm x},t)$. 
For simplicity, we set $\hbar=1$ throughout this paper.

This action is invariant under the global phase transformation 
\be
\psi (x) \rightarrow e^{i \eta} \psi (x)\qquad \mbox{and}\qquad
\psi^\dag (x) \rightarrow e^{-i \eta} \psi^\dag (x) \, , \label{phase}
\end{equation}
where $\eta$ is an arbitrary constant phase.
The Bose-Einstein transition can be regarded as the spontaneous
symmetry breaking associated with this phase transformation.

Usually, the action (\ref{S}) is the basis of the quantization of
trapped BEC systems.
However, in analogy to the Ising model, 
we add the following artificial symmetry breaking term 
to Eq.~(\ref{S}):
\be
\Delta S[\psi,\psi^{\dagger}] = 
\varepsilon {\bar \epsilon} \int\! d\, t\, d^3 x \, 
\left[ e^{-i \theta} v({\bm x}) \psi (x)
+ e^{i \theta} v({\bm x}) \psi^{\dag} (x) \right]. \label{actione}
\end{equation}
Here, ${\bar \epsilon}$ denotes a typical energy scale of the system
determined by trapping frequencies. 
The infinitesimal parameter $\varepsilon$ is taken to be vanishing at
end of the calculation. 
This term is necessary to control the infrared divergence \cite{OY}. 
The order parameter $e^{i \theta} v ({\bm x})$ is determined by
a consistent procedure.

Now we consider canonical formalism. 
The quantized field satisfies
the CCRs
\begin{eqnarray}
\left[\hat{\psi} ({\bm x}, t), \hat{\psi}^\dag ({\bm x'}, t) \right] & = &
  \delta^{(3)} ({\bm x} - {\bm x'}) \, , \label{CCRpsi1} \\
\left[ \hat{\psi} ({\bm x}, t), \hat{\psi} ({\bm x'}, t) \right] & = & 
\left[ \hat{\psi}^\dag ({\bm x}, t), \hat{\psi}^\dag ({\bm x'}, t)
\right] = 0 \, , \label{CCRpsi2}
\end{eqnarray}
where the symbol $\hat{~}$ indicates that quantities with it
are regarded as operators.  
Corresponding to the breakdown of the global phase symmetry,
we divide the original field $\hat{\psi}(x)$ into the classical and
quantum parts, as
\begin{eqnarray}
\hat{\psi}(x) = e^{i\theta} v ({\bm x}) + e^{i\theta} \hat{\varphi} (x) 
\, . 
\label{psi}
\end{eqnarray}
Here, the time-independent real function $v({\bm x})$ is a condensate
field that breaks the global phase symmetry defined by
Eq.~(\ref{phase}). 
Thus, $v (\bm{x})$ is the order parameter of the BEC transition.
It should be noted that $|v (\bm{x})|^2$ gives the number density of
the condensate particles. 
It is assumed that the real number $\theta$ is independent
of both time and space, 
corresponding to the situation without vortices.\footnote{When we 
consider a vortex, the phase depends on the space coordinates.}
In this case, without loss of generality, we can set $\theta=0$.
From the CCRs of $\hat{\psi}$ and $\hat{\psi}^\dag$, (\ref{CCRpsi1}) and
(\ref{CCRpsi2}), the CCRs of $\hat{\varphi}$ and $\hat{\varphi}^\dag$
are given by 
\begin{eqnarray}
 \left[\hat{\varphi} ({\bm x}, t), \hat{\varphi}^\dag ({\bm x'}, t)
 \right] & = & \delta^{(3)} ({\bm x} - {\bm x'}) \, , \label{CCRvphi1} \\ 
\left[ \hat{\varphi} ({\bm x}, t), \hat{\varphi} ({\bm x'}, t) \right] &
 = & \left[ \hat{\varphi}^\dag ({\bm x}, t), \hat{\varphi}^\dag ({\bm
 x'}, t) \right] = 0 \, . \label{CCRvphi2}
\end{eqnarray}

Calculating the Hamiltonian corresponding to the total action $S+\Delta
S$, and substituting Eq.~(\ref{psi}) into it, we obtain
\be
\hat{H} = \hat{H}_0 + \hat{H}_{\rm I} \, , \label{Heps}
\end{equation}
where
\ba
\hat{H}_{0}
&=& \int d^{3} \bm{x}
\,\Bigl[
\hat{\varphi}^{\dag}(K+V - \mu)\hat{\varphi} +
\frac{gv^{2}}{2}(4\hat{\varphi}^{\dag} \hat{\varphi} +
\hat{\varphi}^{2} + \hat{\varphi}^{\dag 2}) \Bigr],  \\
\hat{H}_{\mathrm{I}}
&=& \int d^{3} \bm{x}\,\Bigl[
 v (K+V-\mu + gv^{2} - \varepsilon \bar{\epsilon}) \hat{\varphi} 
 + \hat{\varphi}^{\dag} (K+V-\mu +gv^{2} -\varepsilon
 \bar{\epsilon}) v \nonumber \\ 
&& \hspace{1cm} {}+gv ( \hat{\varphi}^{\dag} \hat{\varphi}^{\dag}
\hat{\varphi} +  \hat{\varphi}^{\dag} \hat{\varphi} \hat{\varphi}) +
\frac{g}{2}\hat{\varphi}^{\dag} \hat{\varphi}^{\dag} \hat{\varphi}
\hat{\varphi} 
\Bigr].
\ea
Here, we have dropped the constant term independent of $\hat{\varphi}$
 and $\hat{\varphi}^{\dag}$.

Now we adopt the Hartree-Fock-Bogoliubov approximation, in which the third-
and fourth-order terms of $\hat{\varphi}$ and $\hat{\varphi}^{\dag}$ are
replaced as follows: 
\begin{subequations}
\ba
\hat{\varphi}^{\dag} \hat{\varphi}^{\dag} \hat{\varphi}
 &\longrightarrow&
 \langle \hat{\varphi}^{\dag} \hat{\varphi}^{\dag} \rangle \hat{\varphi}
 + 2\hat{\varphi}^{\dag} \langle \hat{\varphi}^{\dag} \hat{\varphi}
 \rangle, \\ 
\hat{\varphi}^{\dag} \hat{\varphi} \hat{\varphi}
 &\longrightarrow&
 2\langle \hat{\varphi}^{\dag} \hat{\varphi} \rangle \hat{\varphi}
 + \hat{\varphi}^{\dag}  \langle \hat{\varphi} \hat{\varphi} \rangle, \\
\hat{\varphi}^{\dag} \hat{\varphi}^{\dag} \hat{\varphi} \hat{\varphi}
 &\longrightarrow&
 4\langle \hat{\varphi}^{\dag} \hat{\varphi} \rangle
 \hat{\varphi}^{\dag} \hat{\varphi} + \langle \hat{\varphi}^{\dag2}
 \rangle \hat{\varphi}^2 + \langle \hat{\varphi}^2 \rangle
 \hat{\varphi}^{\dag2}. 
\ea
\end{subequations}
Here, $\langle~\rangle$ represents the thermal expectation value, which
will be determined self-consistently using the mean-field approximated
Hamiltonian $\hat{H}_{\rm MFA}$,
and which is defined through the relation 
$\langle \hat{O} \rangle = {\rm Tr}[e^{-\beta \hat{H}_{\rm MFA}}
\hat{O}]/Z$ with $Z = {\rm Tr}[e^{-\beta \hat{H}_{\rm MFA}}]$, where
$\hat{O}$ is any operator. 
We further employ the Popov approximation
and drop the terms
proportional to  $\langle \hat{\varphi}^2 \rangle$ and $\langle
\hat{\varphi}^{\dag2} \rangle$ \cite{Popov}
in order to ensure a gapless spectrum.\footnote
{
For a homogeneous system, it has been proven that in the 
thermodynamic limit,
if one performs a second-order perturbative calculation, 
the Hugenholtz--Pines (HP) theorem\cite{HP}, which ensures the 
global gauge invariance, 
holds\cite{Popov2}. 
There is yet no similar proof for a trapped system. 
Recently, it has
been shown\cite{Enomoto} that the HP theorem holds for
the loop expansion in the case of a finite volume
system without a trapping potential but with periodic boundary conditions
even without taking the thermodynamic limit.
} 
Finally, the Hamiltonian is reduced to 
\ba
\hat{H}_{\mathrm{MFA}}
&=&\int d^3 \bm{x}\,\Bigl[
\hat{\varphi}^{\dagger}(K +V -\mu
+ 2gv^{2} + 2g\langle \hat{\varphi}^{\dagger}\hat{\varphi}
\rangle)\hat{\varphi} + \frac{g}{2} \hat{\varphi}^{2} v^2 
+\frac{g}{2} \hat{\varphi}^{\dag 2} v^2  \nonumber \\
&& {}+ v (K + V -\mu + gv^2 -\varepsilon \bar{\epsilon}
+ 2g\langle \hat{\varphi}^{\dagger}\hat{\varphi} \rangle )\hat{\varphi}
\nonumber \\ 
&& {}+ \hat{\varphi}^{\dagger}(K + V -\mu
+ gv^2 -\varepsilon \bar{\epsilon}
+ 2g\langle \hat{\varphi}^{\dagger}\hat{\varphi} \rangle )v 
\Bigr]. \label{eqn:ha}
\ea
The terms first-order in $\hat{\varphi}$ and $\hat{\varphi}^{\dagger}$
should vanish. 
Thus, the condensate $v$ satisfies 
\begin{equation}
(K+V-\mu + gv^{2}-\varepsilon\bar{\epsilon}
 + 2g\langle \hat{\varphi}^{\dag}\hat{\varphi}\rangle)v
= 0. \label{eqn:v0}
\end{equation}
This equation is identical to the GP equation \cite{GP} in the limit
$\epsilon \rightarrow 0$.

Because the von Neumann theorem does not hold in QFT, there are many Fock
spaces that are orthogonal to each other \cite{OYVac}, and we need to
choose a physical Fock space among them. 
Here, we adopt the procedure proposed by two of the present authors \cite{OY}, 
in which the physical Fock space is defined by the vacuum of 
annihilation operators defined by diagonalizing $\hat{H}_{\mathrm{MFA}}$
using the GBT\cite{MOY,OY04,OY05}.

First, we prepare an orthonormal set $\{w_n(\bm{x})\}$ obtained
by solving the differential equation
\begin{eqnarray}
\left(K + V -\mu
+ gv^2
+ 2g\langle \hat{\varphi}^{\dagger}\hat{\varphi} \rangle
\right)w_{n}(\bm{x}) = (\epsilon_{n} + \varepsilon
\bar{\epsilon})w_{n}(\bm{x}).\label{eqn:wn} 
\end{eqnarray}
The orthonormal and completeness conditions of this set are
\begin{subequations}
\begin{eqnarray}
\int d^3\bm{x}\, w_{n}(\bm{x}) w_{n'}(\bm{x}) = \delta_{nn'},
 \label{eqn:wnON}\\ 
\sum_{n=0}^{\infty} w_{n}(\bm{x}) w_{n}(\bm{x}') = \delta^{(3)}(\bm{x}-
 \bm{x}') \, . \label{eqn:wncomp}
\end{eqnarray}
\end{subequations}
Then, 
the field operators $\hat{\varphi}(x)$ and $\hat{\varphi}^{\dag}(x)$  are expanded as
\begin{subequations}
\label{eqn:phiexpCS}
\ba
\hat{\varphi}({\bm x}) &=& \sum_{n=0}^{\infty}
 [\hat{b}_n w_{Cn}({\bm x})-\hat{b}_n^\dag w_{Sn}({\bm x})] \, , \\
\hat{\varphi}^\dag({\bm x}) &=& \sum_{n=0}^{\infty}
[ \hat{b}_n^\dag w_{Cn}({\bm x})-\hat{b}_n w_{Sn}({\bm x}) ] \, , 
\ea
\end{subequations}
where 
\ba
w_{Cn}({\bm x}) &=& \sum_{m=0}^{\infty} C_{nm}w_m({\bm x}), \\
w_{Sn}({\bm x}) &=& \sum_{m=0}^{\infty} S_{nm}w_m({\bm x}).
\ea
The matrices $C$ and $S$ are introduced in association with the GBT that
diagonalizes $\hat{H}_{\mathrm{MFA}}$. 
(See Appendix A for details.)
Note that $C$ and $S$ are singular in the limit $\epsilon
\rightarrow 0$ and give rise to an infrared divergence.

Finally, the Hamiltonian $\hat{H}_{\mathrm{MFA}}$ is diagonalized as follows:
\begin{eqnarray}
\hat{H}_{\mathrm{MFA}} = \sum_{n=0}^{\infty} 
E_n \hat{b}^{\dag}_n \hat{b}_n + \mathrm{const} \, , \label{eqn:diagonalized}
\end{eqnarray}
where the quasiparticle energies $E_n$ depend on 
$\{w_n(\bm{x})\}$, $v(\bm{x})$ and
$\{\epsilon_n\}$, and its expression is given in Appendix A. 
It should be noted that the energies $E_n$ are finite even after taking 
the limit of $\epsilon \rightarrow 0$.
Finally, the physical vacuum is defined by 
\begin{eqnarray}
\hat{b}_n |0\rangle = 0.
\end{eqnarray}

The creation and annihilation operators satisfy the bosonic commutation
relations
\begin{eqnarray}
[\hat{b}_n, \hat{b}^{\dag}_{n'}] = \delta_{nn'}, 
\end{eqnarray}
with all other commutators vanishing.
Thus, the quantization scheme explained here 
satisfies the CCRs (\ref{CCRvphi1}) and (\ref{CCRvphi2}) precisely,
in contrast to the Bogoliubov prescription. 
This is because the mode associated with $E_0$ is not included 
in the field expansion in the Bogoliubov prescription.

\section{Dynamical treatment of a system under an external
perturbation}\label{sec-dynamical}

To this point, we have studied the quantization scheme in a static system.
In this section, we discuss the dynamical behavior of a BEC.

To investigate the response to an external field, 
we introduce the time-dependent external perturbation
\begin{eqnarray}
\hat{H}_{\rm ex}(t)=\int d^3\bm{x}\, 
\hat{\psi}^{\dag}(x) \delta V(\bm{x},t) \hat{\psi}(x), \label{eqn:EF}
\end{eqnarray}
where
\begin{eqnarray}
\delta V(\bm{x},t)= \frac{m}{2}\sum^{3}_{i=1}\delta \omega^2_{i}(t) x^2_{i}.
\end{eqnarray}
Then, the total trapping potential is given by 
\begin{eqnarray}
V_{\mathrm{ex}}({\bm x},t)\equiv V({\bm x}) + \delta V(\bm{x},t).
\end{eqnarray}
The Heisenberg equation of motion with a time-dependent perturbation is given by 
\begin{eqnarray}
i \frac{\partial}{\partial t}\hat{\psi}_{\mathrm{ex}}
= (K +
V_{\mathrm{ex}}({\bm x},t)
-\mu-\varepsilon \bar{\epsilon} 
+g\hat{\psi}_{\mathrm{ex}}^{\dagger}
\hat{\psi}_{\mathrm{ex}})\hat{\psi}_{\mathrm{ex}},  \label{eqn:TDGP}
\end{eqnarray}
where $\hat{\psi}_{\mathrm{ex}}$ denotes the boson operator subject to the
perturbation. 
As in the preceding section, we separate $\hat{\psi}_{\mathrm{ex}}$
 into two terms as follows:
\begin{eqnarray}
\hat{\psi}_{\mathrm{ex}}(x) = v_{\mathrm{ex}}({\bm x},t) + \hat{\varphi}(x).
\end{eqnarray}
We further employ the Hartree-Fock-Bogoliubov-Popov (HFBP) approximation.
Then the GP equation becomes
\begin{eqnarray}
i  \frac{\partial}{\partial t}v_{\mathrm{ex}}
=
(K +V_{\mathrm{ex}}
-\mu-\varepsilon \bar{\epsilon} +g|v_{\mathrm{ex}}|^2+2g\langle 
\hat{\varphi}^{\dag} \hat{\varphi} \rangle_{\mathrm{ex}}) v_{\mathrm{ex}},  \label{vexeq}
\end{eqnarray}
where $|v_{\mathrm{ex}}|^2$ is the number distribution of the condensate particles. 
Now, the order parameter has an explicit time dependence, in contrast to 
the static case.

However, Eq. (\ref{vexeq}) is still difficult to solve, because 
we must redefine a physical Fock space 
with the time evolution of the condensate.
If we are interested in low temperature and/or weak coupling systems in which 
fluctuations are expected to be small, however, 
we can further linearize the equation about the equilibrium values.
Thus, the condensate field and the number density of the excited particles, 
$\langle \hat{\varphi}^{\dag} \hat{\varphi} \rangle_{\mathrm{ex}}$, 
are expanded about the static quantities defined in the preceding section,
\ba
v_{\mathrm{ex}}(x) &=& v (\bm{x}) + \delta v(x), \\
\langle \hat{\varphi}^{\dag} \hat{\varphi} \rangle_{\mathrm{ex}} &=& \langle \hat{\varphi}^{\dag} \hat{\varphi} \rangle + \delta \langle \hat{\varphi}^{\dag} \hat{\varphi} \rangle. 
\ea 
Keeping the first-order terms $ \delta V(x)$,
$\delta v(x)$ and $\delta \langle \hat{\varphi}^{\dag} \hat{\varphi}
\rangle$ and ignoring the higher-order terms,
we obtain the linearized equation.
It should be noted that the linear approximation with the HFBP approximation can be 
regarded as the time-dependent linearized Hartree approximation, and 
this approximation is known to be equivalent to the random phase approximation (RPA).
Thus, this approach contains fluctuations beyond the mean-field approximation.
To obtain further simplicity, we may ignore the term proportional to 
$\delta\langle \hat{\varphi}^{\dagger}\hat{\varphi} \rangle$, 
because the quantity 
$\langle \hat{\varphi}^{\dagger}\hat{\varphi} \rangle$
is already very small. 
Finally, we obtain
\ba
i \frac{\partial}{\partial t}\delta v(x)
=
\left(
K +
V-\mu
+ 2gv^2
+ 2g\langle \hat{\varphi}^{\dagger}\hat{\varphi} \rangle
\right)\delta v (x)
+ gv^2 \delta v^{*}(x)
+\delta V(x)
v \, . 
\label{eqn:dv}
\ea
Note that the phase of the condensate field is time dependent.
Thus, this equation can in principle describe vortex formation.

The effect of the zero-mode is included 
in $\langle \hat{\varphi}^{\dagger}\hat{\varphi} \rangle$. 
Thus, it possesses infrared divergence, because of the singularity of 
$S$ and $C$\cite{MOY,OY04,OY05}.
To avoid this difficulty, we employ the renormalization criterion according
to which
all terms proportional to $u^2_0 ({\bm x})$ are ignored.
It should be noted that we still keep the terms proportional to 
$u_l ({\bm x}) u_0 ({\bm x})$ ($l \neq 0$) in the limit of $\epsilon
\rightarrow 0$. 
Thus, the nontrivial effect attributed to the zero-mode is still taken
into account.

\section{Numerical study}\label{sec-Numerical}

In this section, 
we report the results of our numerical study of Eq. (\ref{eqn:dv}).
In this study, we chose the parameter values so as to reproduce 
the experimental situation\cite{Jin}.

\subsection{Thermal particle density} \label{subsec:ncprofile}

According to Eq. (\ref{eqn:dv}), the time evolution 
of the condensate is affected by the thermal non-condensate 
particle density $\langle \hat{\varphi}^{\dag} \hat{\varphi} \rangle$.
First, 
we study the effect of the  zero-mode on the non-condensate particle.
For simplicity, we treat a spatially one-dimensional system 
and consider the 
BEC of rubidium atoms, $^{87}$Rb, whose mass is 
$m=1.42 \times 10^{-25} \mathrm{kg}$, and set 
the frequency of the trapping potential as $\omega_1=200 \times 2 \pi \mathrm{Hz}$. 
Here, we set the parameter values as $T/T_c = 7.3 \times 10^{-2}$, $g=0.02$
in oscillator units (HO units), 
with the condensed atom number $N_{\mathrm{c}} = 1000$. 
Here, $T_c$ is the transition temperature of an ideal Bose gas in a 
one-dimensional system, whose value is
$T_c = \frac{\omega_1}{k_{\mathrm{B}}} \frac{N}{\ln 2 N} = 8.3 \times 10^{-6} \mathrm{K}$, 
according to the estimation given in Ref. \citen{Kett_Dru},
where $k_{\mathrm{B}}$ is the Boltzmann constant and $N$ is the total
number of atoms, which in our calculations is $N=1068$.
The numerical calculations were carried out in the following steps.
First, we diagonalized the
mean field Hamiltonian $\hat{H}_{\mathrm{MFA}}$(\ref{eqn:ha})
under the condition (\ref{eqn:v0}) to define the physical Fock space. 
Accordingly, we estimated the expectation value 
$\langle \hat{\varphi}^{\dagger}\hat{\varphi} \rangle$.

\begin{figure}[t]
\begin{center}
\rotatebox[origin=c]{270}{
\includegraphics[width=6cm,height=9cm]{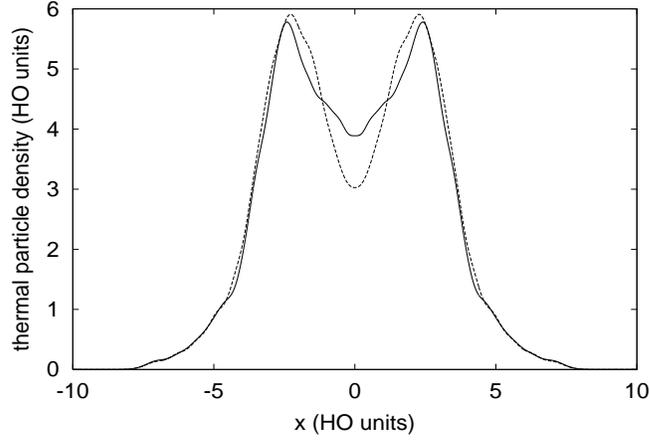}
}
\end{center}
\vspace*{-1.5cm}
\caption{
Thermal particle densities of the one-dimensional system, 
with the parameter values $T/T_c = 7.3 \times 10^{-2}$, $g=0.02$ (HO units). 
The solid curve represents the result with the zero-mode, 
while the dotted curve represents the result without it. 
}
\label{fig:ncprofile.ps}
\end{figure}

Figure \ref{fig:ncprofile.ps} plots this thermal particle density, 
calculated both with and without the zero-mode.
The solid and dashed curves represent the thermal particle density 
with and without the zero-mode, respectively.
One can clearly see that the density of the thermal particles increases 
around $x=0$, because of the zero-mode.
This difference is, however, very small compared to the 
condensate number density, 
because the thermal particle density is about two orders smaller 
than the condensate particle density in our calculation. 
Thus, it is still not clear whether the zero-mode can significantly affect 
the dynamical response of the condensate field.

\subsection{Time evolution of the condensate} \label{subsec:time_evolu}
The actual experiment was carried out in a cylindrically symmetric system.
However, we assume spherical symmetry here for calculational simplicity. 
Then, the static trap potential with frequency
$\omega\equiv\omega_i \, (i=1,2,3)$ and the corresponding 
external perturbation are given by 
\begin{eqnarray}
V(r) &=& \frac{1}{2} m \omega^2 r^2 , \\
\delta V (r,t) &=& \frac{1}{2} m \delta \omega^2(t) r^2 ,
\end{eqnarray}
respectively.
In conformity with the experiment\cite{Jin}, we employed the
following time-dependent frequency:
\begin{eqnarray}
\delta \omega^2(t) = \omega^2 A \sin \Omega t.
\end{eqnarray}
The quantities $A$ and $\Omega$ are the driving amplitude and the 
driving frequency, respectively.

The numerical calculation was carried out in the following steps.
First, we estimated the expectation value 
$\langle \hat{\varphi}^{\dagger}\hat{\varphi} \rangle$, 
using the procedure described in the previous subsection, 
and 
solved Eq. (\ref{eqn:dv}).
 The parameter values were chosen so as to reproduce 
the trapped atoms of $^{87}$Rb, where 
$m=1.42 \times 10^{-25} \mathrm{kg}$, 
$\omega=200 \times 2 \pi \mathrm{Hz}$ and  
$g=0.03$ in HO units and condensed atom number is $N_{\mathrm{c}} = 1000$.
The temperature $T$, in units of $T_{\mathrm{c}}$, the driving amplitude $A$, and 
the driving frequency $\Omega$ are the control parameters in the experiment, 
where $T_{\mathrm{c}} = 0.94 \frac{\omega}{k_{\mathrm{B}}} N^{1/3}$ is the critical temperature of the ideal Bose gas confined in a three-dimensional spherically symmetric trap\cite{Dalfovo}. 
The value of $T_{\mathrm{c}}$ is approximately 100 nK in this calculation.

\begin{figure}[t]
\begin{center}
\rotatebox[origin=c]{270}{
\includegraphics[width=6cm,height=9cm]{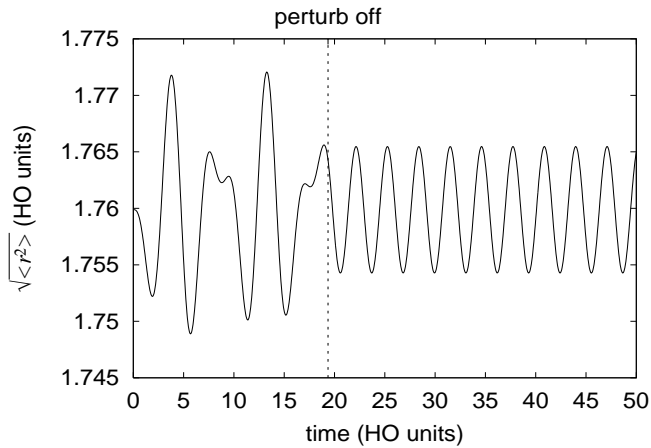}
}
\end{center}
\vspace*{-1.5cm}
\caption{Time evolution of the condensate with the parameter values
$T/T_{c}=0.11$, $A=0.01$, $\Omega=1.3$ (HO units). 
The expression ``perturb off'' refers to the time at which the perturbation is turned off. 
After that time, the condensate oscillates sinusoidally
with an amplitude of 0.3 \% with respect to the initial size. }
\label{fig:omega13.eps}
\end{figure}

In Fig. \ref{fig:omega13.eps}, we plot the time evolution of the 
size of the condensate, defined by 
\begin{eqnarray}
\sqrt{\langle r^2 \rangle} = \sqrt{\frac{\int r^2 | v_{\mathrm{ex}} |^2\,d^3 \bm{x}}{\int |v_{\mathrm{ex}}|^2\,d^3 \bm{x}}} \, \, \, .
\end{eqnarray}
The temperature, the driving amplitude and the driving frequency 
are given by 
$T/T_c=0.11$, $A=0.01$ and $\Omega=1.3\,\text{\,(HO units)}$, 
respectively.
Under the external perturbation, we observe beat-like behavior.
The external perturbation was turned off at  
about 15.4 ms, which is a typical value in the corresponding 
experiment \cite{Jin}. 
Then, the frequency of the oscillation of the condensate is 2.014 (HO units).
\begin{figure}[t]
\begin{center}
\rotatebox[origin=c]{270}{
  \includegraphics[width=6cm,height=9cm]{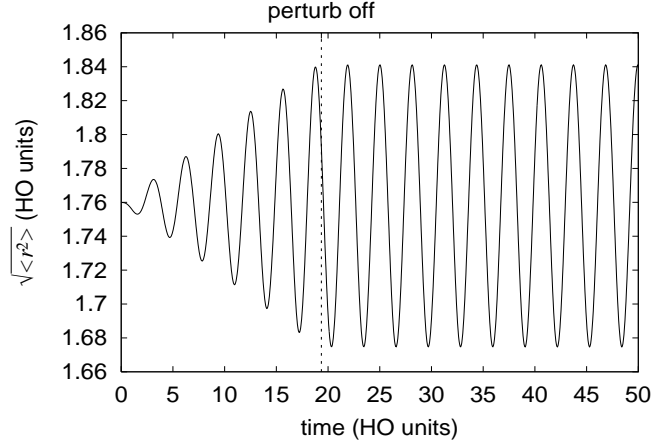}
}
\end{center}
\vspace{-1.5cm}
\caption{Time evolution of the condensate with the parameter values 
$T/T_{c}=0.11$, $A=0.01$, $\Omega=2.0$ (HO units). 
The amplitude of the oscillation is 4.5 \% with respect to the initial size. }
\label{fig:omega2.eps}
\end{figure}
This frequency characterizes the dynamical behavior of 
the condensate.
We plot the numerical result with a different driving frequency, 
$\Omega=2.0\,\text{(HO units)}$, in Fig. \ref{fig:omega2.eps}.
After turning off the external perturbation, 
the frequency of the oscillation of the condensate is again 
given by 2.014 (HO units).
It should be noted that
the resonance gives rise to an increase of the amplitude, 
and 
we verified that the amplitude is maximized at $\Omega = 2.014$ (HO units).

The first excitation energy of the quasiparticle 
is $E_1 = 2.014$ (HO units) [see Eq. (\ref{eqn:diagonalized})].
It is thus seen that the frequency of the oscillation is characterized by 
the first excitation energy of the quasiparticle.
This result is consistent with that of the 
linear density response \cite{Minguzzi}.

\subsection{Temperature dependence of the response frequency and the 
effect of the zero-mode} \label{subsec:Tdep}

In the preceding section, we showed that it is possible to extract 
information concerning the first excitation energy of the 
quasiparticle through the frequency of the oscillation.
In this section, we investigate the effect of the zero-mode on the 
frequency.

In Table \ref{table:1}, 
the temperature dependence of the frequency is described for two cases, 
those in which the frequencies are calculated with the zero-mode 
and those without the zero-mode.
The driving amplitude and the driving frequency are given by 
$A=0.01$ and $\Omega=2.0$ (HO units), respectively.
The table lists the temperature, the frequency with the zero-mode, 
that without the zero-mode, and 
the difference between these two frequencies.
It can be seen that 
the effect of the zero-mode on the response frequency is very small.
This is consistent with the fact that the properties of a trapped BEC at low temperature 
are described in the Bogoliubov prescription, although it violates the CCRs.
It is also seen that the frequency decreases as the temperature is increased.
It follows that the first excitation energy is lowered by the influence of 
the medium effect.
Thus, the energy shift 
due to the zero-mode becomes important as the temperature is increased.
However, the magnitude of this shift is still very small, and the effect of the zero-mode is negligible 
for this set of parameter values.

   \begin{table}
     \caption{Temperature dependence of response frequency, calculated with and without the zero-mode.}
     \label{table:1}
     \begin{center}
       \begin{tabular}{cccc} \hline \hline
                & \multicolumn{2}{c}{response frequency }  \\ 
        $T/T_c$ & with zero-mode   & without zero-mode & difference \\ \hline
         0.106 &  2.01537  &  2.01533 & 0.00004\\
         0.312 &  2.11152  &  2.01139 & 0.00013\\ 
         0.507 &  2.00710  &  2.00688 & 0.00022\\ \hline
     \end{tabular}
     \end{center}
     \end{table}

\subsection{Coupling dependence of the first excitation energy and the 
effect of the zero-mode}

In the preceding subsection, we showed that the effect of the zero-mode in the experiment reported in Ref. \citen{Jin} 
is quite small and does not affect the properties of the trapped BEC system.
The merit of the trapped BEC experiment is that we can change the magnitude of the coupling constant 
by using the Feshbach resonance\cite{Feshbach}.
Thus, the effect of the zero-mode may be experimentally observed with 
other set of parameter values.
In this section, we investigate the coupling constant dependence of the zero-mode. 

We again assume an one-dimensional system and 
consider a BEC of rubidium atoms, $^{87}$Rb. 
The parameter values are the same as in \S \S \ref{subsec:ncprofile}. 
Here, we simply calculate the first excitation energy,
which corresponds to the dipole mode. 
This is an ideal one-dimensional calculation, and 
it does not correspond directly to the experiments. 

   \begin{wraptable}{l}{\halftext}
     \caption{Temperature and coupling constant dependences of the 
first excitation energy including the zero-mode.}
     \label{table:3}
     \begin{center}
       \begin{tabular}{cccc} \hline \hline
                & \multicolumn{3}{c}{first excitation energy }  \\ 
                       $T/T_c$   & g=0.01   & g=0.015 & g=0.02 \\ \hline
  $1.5 \times 10^{-2}$  &  1.0033  &  1.0454 & 1.0749\\
  $3.7 \times 10^{-2}$  &  1.0685  &  1.1243 & 1.1908\\
  $7.3 \times 10^{-2}$  &  1.143  &  1.243 & 1.350 \\ \hline
       \end{tabular}
     \end{center}
   \end{wraptable}

   \begin{wraptable}{l}{\halftext}
     \caption{Temperature and coupling constant dependences of the subtraction 
of the frequency without the zero-mode 
     from that that with the zero-mode.}
     \label{table:4}
     \begin{center}
       \begin{tabular}{cccc} \hline \hline
                & \multicolumn{3}{c}{difference }  \\ 
        $T/T_c$   & g=0.01   & g=0.015 & g=0.02 \\ \hline
 $1.5 \times 10^{-2}$  &  0.0007  &  0.0020 & 0.0037\\
 $3.7 \times 10^{-2}$  &  0.0022  &  0.0046 & 0.0088\\
 $7.3 \times 10^{-2}$  &  0.004  &  0.008 & 0.013\\ \hline
       \end{tabular}
     \end{center}
   \end{wraptable}

Table \ref{table:3} lists the temperature and coupling constant 
dependences of the first excitation energy including the zero-mode. 
We can see that this energy becomes larger as we increase not only the
temperature but also the coupling constant.
The difference\footnote
{The definition of this quantity is the same as that given in Table \ref{table:1}.}
is listedd in Table \ref{table:4}.
\footnote{The coupling constant $g$ used here is smaller than $0.03$,
the value used in the calculations 
in the preceding subsections. 
Note that the spatial dimension is 1 in this subsection, while it is
3 in \S \S \ref{subsec:time_evolu} and \ref{subsec:Tdep}.
}
However, 
our numerical results exhibit deviation from  Kohn's theorem\cite{Kohn}, 
according to which 
the energy of the dipole mode should be 1.
This seems to be due to the fact that the dynamics of the non-condensate 
field are not properly taken into account.
As a matter of fact, the deviation becomes larger as the temperature is increased.
Thus, the effect of the zero-mode observed here is not physical.

\vspace{5cm}

   \begin{wraptable}{l}{\halftext}
     \caption{Temperature and coupling constant dependences of the second excitation energy including the zero-mode.}
     \label{table:5}
     \begin{center}
       \begin{tabular}{cccc} \hline \hline
                & \multicolumn{3}{c}{second excitation energy }  \\ 
                       $T/T_c$   & g=0.01   & g=0.015 & g=0.02 \\ \hline
  $1.5 \times 10^{-2}$  &  1.7874  &  1.7873   & 1.7974 \\
  $3.7 \times 10^{-2}$  &  1.8199  &  1.8409   & 1.8744\\
  $7.3 \times 10^{-2}$  &  1.875   &  1.928    & 1.991\\ \hline
       \end{tabular}
     \end{center}
   \end{wraptable}

   \begin{wraptable}{l}{\halftext}
     \caption{Temperature and coupling constant dependences of the difference (second excitation energy).}
     \label{table:6}
     \begin{center}
       \begin{tabular}{cccc} \hline \hline
                & \multicolumn{3}{c}{difference }  \\ 
        $T/T_c$   & g=0.01   & g=0.015 & g=0.02 \\ \hline
 $1.5 \times 10^{-2}$  &  0.0007  &  0.0018  & 0.0030 \\
 $3.7 \times 10^{-2}$  &  0.002  &   0.004 & 0.008 \\
 $7.3 \times 10^{-2}$  &  0.004  &  0.007 & 0.012\\ \hline
       \end{tabular}
     \end{center}
   \end{wraptable}

As we have discussed, the first excitation energy should be 1 
for all temperatures and interactions, as asserted by Kohn's theorem.
In order to determine the effect of the zero-mode, we should calculate the 
energy shift of the second excitation energy.
Table \ref{table:5} lists the temperature and coupling constant 
dependence of the second excitation energy including the zero-mode, 
and the difference is listed in Table \ref{table:6}. 
It is seen that the effect of the zero-mode is enhanced as the temperature and the coupling 
constant are increased.
However, by comparing with Table \ref{table:4}, 
we see that 
the effect of the zero-mode on the second excitation energy is of almost the same order as that 
on the first excitation energy.
Thus, there is a possibility that this effect vanishes when we 
make corrections in order
to satisfy Kohn's theorem.
In short, the effect of the zero-mode in the one-dimensional system is 
not yet clear.

\section{Concluding remarks}\label{sec-summary}

We have formulated the dynamical response of a trapped BEC 
including the zero-mode 
at low temperature, using the HFBP and linear approximations. 
The zero-mode appears as a NG mode associated with the
global phase symmetry breaking. 
To take the zero-mode into account, we introduced the generalized
Bogoliubov transformation following Ref.~\citen{OY}. 
We numerically calculated the dynamical response of the condensate 
by applying a time-dependent perturbation in the 
linear response approximation.
Thus, the fluctuations from the mean-field approximation are partially
taken into account.
The frequency of the condensate is determined by the first excitation
energy of the quasiparticle.
We observed that the presence of the zero-mode increases the frequency 
and hence the first excitation energy.
With the parameter values used in the experiment\cite{Jin}, the effect of the zero-mode 
on the first excitation energy is very small.
Thus, we conclude that the properties of the trapped BEC in this region can be 
described by the Boboliubov prescription, 
although it leads to an inconsistent description in quantum field theory.

To determine the temperature and coupling constant dependences, 
we studied the first and second excitation energies in a one-dimensional system.
From Kohn's theorem, it can be proved that the first excitation energy should be 1 for any temperature and interaction.
Our results, however, exhibit deviation from 1.
This deviation appears to be due to the inadequate approximation used for 
the non-condensate field, because the deviation becomes larger as the temperature 
is increased.
The second excitation mode is also affected by the zero-mode.
The magnitude of the energy shift is, however, almost the same  
as that of the first excitation energy.
Thus, it is not clear whether the observed effect comes from the 
effect of the zero-mode or from the inadequate approximation violating Kohn's theorem.

To improve our results, we should take account of 
the fluctuations of the non-condensate field, 
which is dropped in Eq. (\ref{eqn:dv}).
As a matter of fact, it has been reported that 
the collective dynamics of the non-condensate field play an important role\cite{ref:Pita}.
In our calculation, 
if we take the fluctuations of the non-condensate field into account, 
the approximation is equivalent to the time-dependent linear Hartree approximation, 
that is, the RPA, and we can expect that a collective mode like the plasmon 
in an electron gas will be included.
However, the translational invariance is broken by the trapped potential 
in this system.
Thus, the analysis of the collective mode is not simple, 
and we must introduce 
several approximations to simplify calculations\cite{Morgan2,Giorgini}.
The study of the effect of the fluctuations of the non-condensate field is 
a future problem.

In this paper, we have concentrated on elucidating the effect
of the zero-mode on the first excitation energy.
However, calculations of the temperature dependence of the
condensate fractions and the specific heat are also interesting.

The effect of the gapless mode was also studied in Ref.~\citen{Hut}.
However, that work does not include the zero-mode in the non-condensate field. 
Consequently, the non-condensate field has zero projection on the
condensate, and this property simplifies the calculation.
By contrast, the zero-mode is included in the non-condensate field 
in our formalism.
Thus, the gapless mode investigated in Ref.~\citen{Hut} is different from the zero-mode
investigated here.

\section*{Acknowledgements}
The authors would like to thank Professors I.~Ohba and H.~Nakazato 
for helpful comments and encouragement. 
The authors thank the Yukawa Institute for Theoretical Physics 
at Kyoto University for offering us the opportunity to discuss this work 
during the YITP workshop YITP-W-03-10, ``Thermal
Quantum Field Theories and Their Applications''.
M.M. is supported partially by Fujukai and by a Grant-in-Aid 
for The 21st Century COE Program (Physics of Self-organization Systems) 
at Waseda University.
T.K. acknowledges a fellowship from the FAPESP(04/09794-0).
This work is partially supported by a Grant-in-Aid for Scientific Research
(C) (No.~17540364) from the Japan Society for the Promotion of Science,
and by Grants-in Aid for Young Scientists (B) (No.~17740258) and for
Priority Area Research (B) (No.~13135221) both from the Ministry of
Education, Culture, Sports, Science and Technology, Japan.
M.O. and Y.Y. wish to express their thanks for 
a Waseda University Grant for Special Research Projects.

\appendix
\section{Generalized Bogoliubov Transformation Including the Zero-Mode}

We have introduced the orthonormal 
set $\{ w_n(\bm{x}) \}$ satisfying  Eq.~(\ref{eqn:wn}) with the orthonormal
and completeness conditions (\ref{eqn:wnON}) and (\ref{eqn:wncomp}).
It is easily seen that Eq.~(\ref{eqn:wn}) reduces to the GP
equation for $n=0$ in the limt $\epsilon \rightarrow 0$, 
because we have $\epsilon_{0}=0$. 
The condensate field is given by 
\begin{eqnarray}
v (\bm{x}) = \sqrt{N_\mathrm{c}} w_0 (\bm{x}). 
\end{eqnarray}
Using the orthonormal set, 
we expand $\hat{\varphi}(x)$ and introduce the creation and annihilation operators: 
\begin{eqnarray}
\hat{\varphi}(\bm{x}) = \sum_{n=0}^{\infty} \hat{a}_{n} w_{n}(\bm{x}). \label{eqn:phi}
\end{eqnarray}
The operators $\hat{a}_n$ and $\hat{a}_n^{\dag}$ are subject to the condition
\begin{eqnarray}
[\hat{a}_n, \hat{a}_{n'}^{\dag}] = \delta_{nn'}, 
\end{eqnarray}
with all other commutators vanishing:
$[\hat{a}_n, \hat{a}_{n'}] =[\hat{a}^{\dag}_n, \hat{a}_{n'}^{\dag}] =0$.

Substituting Eq.~(\ref{eqn:phi}) into the mean field Hamiltonian (\ref{eqn:ha}) 
with the condition (\ref{eqn:v0}), 
we obtain
\begin{eqnarray}
\hat{H}_{\mathrm{MFA}}
&=& \sum_{n=0}^{\infty} (\epsilon_{n} + \varepsilon \bar{\epsilon}) \hat{a}_{n}^{\dag} \hat{a}_{n} 
+ \sum_{n,n'=0}^{\infty}
\Bigl[
2U_{nn'}\hat{a}_{n}^{\dag}\hat{a}_{n'}
+ U_{nn'} \hat{a}_{n} \hat{a}_{n'} 
+ U_{nn'}\hat{a}_{n}^{\dag} \hat{a}_{n'}^{\dag}
\Bigr],  
\end{eqnarray}
where
\begin{eqnarray}
U_{nn'}&=&\frac{g}{2}\int d^3\bm{x}\, w_{n}(\bm{x}) w_{n'}(\bm{x})
 v^{2}(\bm{x}). 
\end{eqnarray}

The physical Fock space is defined by the creation and
annihilation operators that diagonalize the mean field Hamiltonian. 
For this purpose, the following operators are introduced: 
\begin{subequations}
\begin{eqnarray}
\hat{q}_n = \sqrt{\frac{1}{2(\varepsilon_n + \varepsilon \bar{\epsilon})}}
(\hat{a}_n + \hat{a}^{\dag}_n), 
\end{eqnarray}
\begin{eqnarray}
\hat{p}_n = -i \sqrt{\frac{\varepsilon_n + \varepsilon \bar{\epsilon}}{2}}
(\hat{a}_n - \hat{a}^{\dag}_n). 
\end{eqnarray}
\end{subequations}
These operators satisfy the CCRs
\begin{eqnarray}
[\hat{q}_n , \hat{p}_n] = i \delta_{nn'}.
\end{eqnarray}
All other commutators vanish.

Then, $\hat{H}_\mathrm{MFA}$ can be rewritten as 
\begin{eqnarray}
\hat{H}_\mathrm{MFA} 
=\sum_{n=0}^{\infty} \frac{1}{2}\hat{p}_n^2 
+\sum_{n,n'=0}^{\infty} \frac{1}{2}\hat{q}_n W_{nn'} \hat{q}_{n'} 
-\sum_{n=0}^{\infty} 
\Bigl[ 
\frac{1}{2}(\varepsilon_n + \varepsilon \bar{\epsilon}) + U_{nn'}
\Bigr]. 
\end{eqnarray}
Here, the matrix $W$ has a block structure, 
\begin{eqnarray}
W = 
\left(
\begin{array}{cc}
4 (\varepsilon \bar{\epsilon}) U_{00} + O (\varepsilon^2) & 
\sqrt{\varepsilon \bar{\epsilon}} {u'}^{\rm T}
+ O (\varepsilon^{\frac{3}{2}}) \\
\sqrt{\varepsilon \bar{\epsilon}} \, u'
+ O (\varepsilon^{\frac{3}{2}}) & 
W' + O (\varepsilon)
\end{array}
\right) \, ,
\label{W}
\end{eqnarray}
with
\begin{subequations}
\begin{eqnarray}
u' & = &
\left(
\begin{array}{c}
4 \sqrt{\epsilon_1} U_{10} \\
4 \sqrt{\epsilon_2} U_{20} \\
 \vdots
\end{array}
\right), \\
 W'_{nm} & = & \epsilon_n^2 \delta_{nm} 
+ 4\sqrt{\epsilon_n} U_{nm} \sqrt{\epsilon_{m}} \, . \label{Wele} \\
&& \qquad \qquad (n, m=1,2,\cdots)  \nn
\end{eqnarray}
\end{subequations}
This symmetric matrix $W$ can be diagonalized by 
using an orthogonal matrix {\cal O} with eigenvalues
 $E_n^2\ (n=0,1,2,\cdots)$: 
\begin{eqnarray}
{\cal O} W {\cal O}^{\rm T} =  E^2 \, , 
\label{diagW}
\end{eqnarray}
with
\begin{eqnarray}
(E)_{nm}=E_n \delta_{nm} \, . \label{Enm}
\end{eqnarray}

The above expressions are expanded with respect to the infinitesimal
parameter $\varepsilon$.  It should be noted that 
the zero-th eigenvalue is given by
\begin{subequations}
\label{EG0}
\ba
E_0 & = & \sqrt{\varepsilon \bar{\epsilon}} \sqrt{\bar{E}}_0
 + O (\varepsilon^{\frac{3}{2}}) \, , \\
\bar{E}_0& \equiv &  4U_{00}- 
u^{'\rm T}  W^{'-1} u' \, ,
\ea
\end{subequations}
and it vanishes when we take the limit $\epsilon \rightarrow 0$, 
but we keep it finite in the quantization procedure.

Then, the orthogonal matrix ${\cal O}$ can be written as 
\begin{eqnarray}
 {\cal O} = 
 \left(
\begin{array}{cc}
1 - \frac{1}{2} (\varepsilon \bar{\epsilon}) 
u^{'\rm T}  W^{'-2} u' + O (\varepsilon^2) &
- \sqrt{\varepsilon \bar{\epsilon}} \, 
u^{'\rm T} W^{'-1}{\cal O'}^{\rm T}
+ O (\varepsilon^{\frac{3}{2}}) \\
\sqrt{\varepsilon \bar{\epsilon}} \, 
 {\cal O'}  W^{'-1} u'
+ O (\varepsilon^{\frac{3}{2}}) &
{\cal O'} + O (\varepsilon)
\end{array}
\right) \, ,
\end{eqnarray}
where ${\cal O'}$ is another orthogonal 
matrix diagonalizing the matrix $W'$,
\begin{eqnarray}
{\cal O'} W' {\cal O}^{'\rm T} = E^{'2} \, , \label{diagW'}
\end{eqnarray}
with the diagonal matrix
\begin{eqnarray}
\left( E^{'2} \right)_{nm} = E_n^2 \delta_{nm} 
+ O (\varepsilon^{\frac{1}{2}}) \, .
\quad (n,m=1,2,\cdots)  \label{E'E}
\end{eqnarray}
Thus the mixing among $\hat{a}_n\ (n=1,2,\cdots)$ is regular 
with respect to $\varepsilon$ ;
only the mixing between $\hat{a}_0$ and other $\hat{a}_n$ gives rise 
to a singularity.

By using the orthogonal matrix ${\cal O}$, 
we can introduce new pairs of canonical operators as
\begin{subequations}
\begin{eqnarray}
\hat{Q}_n = \sum_{m=0}^{\infty} {\cal O}_{nm} \hat{q}_m \, ,
\end{eqnarray}
\ba
\hat{P}_n = \sum_{m=0}^{\infty} {\cal O}_{nm} \hat{p}_m \, ,
\ea
\end{subequations}
with $[\hat{Q}_n, \hat{P}_{n'}] = i \delta_{nn'}$, 
while all other commutators vanish.

Finally, we can diagonalize the mean field Hamiltonian,
\ba
\hat{H}_\mathrm{MFA} 
&=& 
\sum_{n=0}^{\infty} 
\Bigl[
\frac{1}{2}\hat{P}_n^2 
+ \frac{1}{2}\hat{Q}_n^2 
-\frac{1}{2}(\varepsilon_n + \varepsilon \bar{\epsilon}) + U_{nn'}
\Bigr] \nonumber \\
&=& 
\sum_{n=0}^{\infty} 
\Bigl[
E_n \hat{b}^{\dag}_n \hat{b}_n + \frac{1}{2} E_n
-\frac{1}{2}(\varepsilon_n + \varepsilon \bar{\epsilon}) + U_{nn'}
\Bigr]. \label{Hbb}
\ea
In the second equality, the operators $ \hat{Q}_n$ and $\hat{P}_n$ are
related to $\hat{b}_n$ and $\hat{b}^{\dag}_n$ as 
\begin{subequations}
\ba
\hat{Q}_n = \sqrt{\frac{1}{2 E_n}} (\hat{b}_n + \hat{b}^{\dag}_n) \, ,
\ea
\ba
\hat{P}_n = -i \sqrt{\frac{E_n}{2}} (\hat{b}_n + \hat{b}^{\dag}_n) \, .
\ea
\end{subequations}
Thus, the physical vacuum is defined as
\begin{eqnarray}
\hat{b}_n |0\rangle = 0.
\end{eqnarray}

In short, the diagonalization studied to this point is 
realized by introducing the generalized Bogoliubov transformation,
\begin{subequations}
\label{eqn:bn}
\ba
\hat{b}_{n} = \sum_{m=0}^{\infty} \left[C_{nm} \hat{a}_{m} + S_{nm} \hat{a}_{m}^{\dag} \right] \, ,\\
\hat{b}_{n}^{\dag} = \sum_{m=0}^{\infty} \left[C_{nm} \hat{a}_{m}^{\dag} + S_{nm} \hat{a}_{m} \right] \, .
\ea
\end{subequations}
Here the real matrices $C$ and $S$ in the matrix notation 
are given by
\begin{subequations}
\label{CS}
\ba
C & = & \frac{1}{2} \left( 
E^{\frac{1}{2}} {\cal O} \epsilon^{-\frac{1}{2}}
+ E^{-\frac{1}{2}} {\cal O} \epsilon^{\frac{1}{2}}
\right) \, ,
\\
 S & = & \frac{1}{2} \left( 
E^{\frac{1}{2}} {\cal O} \epsilon^{-\frac{1}{2}}
-E^{-\frac{1}{2}} {\cal O} \epsilon^{\frac{1}{2}}
\right) 
\, , 
\ea
\end{subequations}
with $E$ defined by (\ref{Enm}) and 
\begin{subequations}
\label{matrixEGvare}
\ba
\left(\epsilon \right)_{nm}&=&( \epsilon_n 
+ \varepsilon {\bar \epsilon} ) \delta_{nm}
\, .
\ea
\end{subequations}

One can easily check the properties of $C$ and $S$ from Eq. (\ref{CS}),
and we find
\begin{subequations}
\ba
\sum_{m=0}^{\infty} (C_{nm} C_{n'm} - S_{nm} S_{n'm}) = \delta_{nn'} \, , \\
\sum_{m=0}^{\infty} (C_{nm} S_{n'm} - S_{nm} C_{n'm}) = 0 \, .
\ea
\end{subequations}
Thus, the field operators can be expanded in terms
of $\hat{b}$-operators diagonalizing the Hamiltonian $\hat{H}_\mathrm{MFA}$ as
\begin{subequations}
\label{varphib}
\ba
\hat{\varphi}(x) &=& \sum_{n=0}^{\infty}
 [\hat{b}_n e^{- i E_n t} w_{Cn}({\bm x})-\hat{b}_n^\dag e^{i E_n t}
 w_{Sn}({\bm x})] \, , \\ 
\hat{\varphi}^\dag(x) &=& \sum_{n=0}^{\infty}
[ \hat{b}_n^\dag e^{i E_n t} w_{Cn}({\bm x})-\hat{b}_n e^{- i E_n t}
w_{Sn}({\bm x}) ]  \, ,
\ea
\end{subequations}
with
\begin{subequations}
\label{defwcws}
\ba
w_{Cn}({\bm x}) &=& \sum_{m=0}^{\infty} C_{nm}w_m({\bm x}) \, ,\\
w_{Sn}({\bm x}) &=& \sum_{m=0}^{\infty} S_{nm}w_m({\bm x}) \, .
\ea
\end{subequations}

It is emphasized that the canonical commutation relation
\begin{eqnarray}
[ \hat{\varphi} ({\bm x},t), \hat{\varphi}^\dag ({\bm x'},t) ] &=& 
\delta^{(3)} ({\bm x} - {\bm x'}) 
\end{eqnarray}
holds in this quantization scheme. 
This is because we introduced the artificial interaction ($\ref{actione}$) 
so that we could control the infrared divergence, 
as seen in Eqs. (\ref{EG0}), (\ref{eqn:bn}) and (\ref{CS}).


\begin{thebibliography}{99}
%
\bibitem{GP}
E.~P.~Gross, Nuovo Cim. {\bf 20} (1961), 454; 
 J. Math. Phys. {\bf 4} (1963), 195. \\
L.~P.~Pitaevskii, Zh. Eksp. Teor. Fiz. [Sov. Phys. JETP] 
{\bf 40} (1961), 646; Sov. Phys. JETP {\bf 13} (1961), 451.
%
\bibitem{Dalfovo}
F.~Dalfovo, S.~Giorgini, L.~P.~Pitaevskii and S.~Stringari, 
Rev. Mod. Phys. \textbf{71} (1999), 463.
%
\bibitem{Feshbach}
S.~Inouye, M.~R.~Andrews, J.~Stenger, H.-J.~Miesner, D.~M.~Stamper-Kurn 
and W.~Ketterle, Nature {\bf 392} (1998), 151. 
%
\bibitem{Gri1}
A.~Griffin, Phys.~Rev. B {\bf 53} (1996), 9341.
%
\bibitem{Gri2}
D.~A.~W.~Hutchinson, E.~Zaremba and A.~Griffin, 
Phys.\ Rev.\ Lett. {\bf 78} (1997), 1842.
%
\bibitem{Hut}
D.~A.~W.~Hutchinson et al., J.~of~Phys. B {\bf 33} (2000), 3825.
%
\bibitem{Rus}
M.~Rusch et al., Phys.\ Rev.\ Lett. {\bf 85} (2000), 4844.
%
\bibitem{Choi}
J.~Rogel-Salazar et al., J.~of~Phys. B {\bf 34} (2001), 4617.
%
\bibitem{Kru}
V.~I.~Kruglov, M.~K.~Olsen and M.~J.~Collett,
Phys.~Rev.~A {\bf 72} (2005), 033604.
%
\bibitem{JZ}
B.~Jackson and E.~Zaremba,
Phys.~Rev.~Lett. {\bf 88} (2002), 180402.
%
\bibitem{MRHB}
S.~A.~Morgan, M.~Rusch, D.~A.~W.~Hutchinson and K.~Burnett,
Phys.~Rev.~Lett. {\bf 91} (2003), 250403.
%
\bibitem{Morgan1}
S.~A.~Morgan, 
Phys.\ Rev.\ A {\bf 69} (2004), 023609.
%
\bibitem{Morgan2}
S.~A.~Morgan, 
Phys.\ Rev.\ A {\bf 72} (2005), 043609.
%
\bibitem{Jin}
D.~S.~Jin, M.~R.~Matthews, J.~R.~Ensher, C.~E.~Wieman and E.~A.~Cornell, 
Phys.\ Rev.\ Lett. {\bf 78} (1997), 764.
%
\bibitem{ZNK}
E.~Zaremba, T.~Nikuni and A.~Griffin, 
J. Low Tmep. Phys. {\bf 116} (1999), 277.
%
\bibitem{Ramos1}
D.~G.~Barci, E.~S.~Fraga and R.~O.~Ramos, 
Phys.\ Rev.\ Lett. {\bf 85} (2000), 479.
%
\bibitem{Ramos2}
D.~G.~Barci, E.~S.~Fraga, M.~Gleiser and R.~O.~Ramos,
Physica~A {\bf 317} (2003), 535.
%
\bibitem{Ramos3}
J.-L. Kneur, M. B. Pinto and R. O. Ramos,
Phys.\ Rev.\ Lett. {\bf 89} (2002), 210403.
%
\bibitem{Boya}
D.~Boyanovsky et al., Ann.~of~Phys.~{\bf 300} (2002), 1.
%
\bibitem{OY}
M.~Okumura and Y.~Yamanaka, 
Phys. Rev. A \textbf{68} (2003), 13609.
%
\bibitem{OY05}
M.~Okumura and Y.~Yamanaka, Physica A {\bf 348} (2005), 157.
%
\bibitem{OYVac}
M.~Okumura and Y.~Yamanaka, Physica A, to be published. 
%
\bibitem{Lewen}
M.~Lewenstein and L.~You, Phys.\ Rev.\ Lett. {\bf 77} (1996), 3489. 
%
\bibitem{Matsu}
H.~Matsumoto and S.~Sakamoto, Prog. Theor. Phys. {\bf 107} (2002), 679.
%
\bibitem{MOY}
M.~Mine, M~Okumura and Y.~Yamanaka, J. Math. Phys. {\bf 46} (2005), 042307.
%
\bibitem{OY04}
M.~Okumura and Y.~Yamanaka, Prog. Theor. Phys. {\bf 111} (2004), 199.
%
\bibitem{Griffin}
A. Griffin, Phys. Rev. B \textbf{53} (1996), 9341
%
\bibitem{Popov}
V.~N.~Popov, \textit{Functional Integrals and Collective Excitations} (Cambridge University Press, Cambridge, 1987).
%
\bibitem{HP}
N.~Hugenholtz and D.~Pines, Phys. Rev. {\bf 116} (1959), 489.
%
\bibitem{Popov2}
V.~N.~Popov, Sov. Phys. JETP {\bf 20} (1965), 1185.
%
\bibitem{Enomoto}
H.~Enomoto, M.~Okumura and Y.~Yamanaka, 
to be published in Ann.~of~Phys. 
%
\bibitem{Kohn}
W.~Kohn, 
Phys. Rev. \textbf{123} (1961), 1242.
%
\bibitem{Minguzzi}
A.~Minguzzi and M.~P.~Tosi, J.~of~Phys.: Cond. Mat. {\bf 9} (1997), 10211.
%
\bibitem{Kett_Dru}
W. ~Ketterle and N.~J.~van Druten, 
Phys. Rev. A {\bf 54}, (1996), 656.
%
\bibitem{ref:Pita}
L.~Pitaevskii and S.~Stringari, Phys. Rev. Lett. {\bf 81} (1998), 4541.
%
\bibitem{Giorgini}
S.~Giorgini, 
Phys. Rev. A \textbf{61} (2000), 063615.
%
\end{thebibliography}
\end{document}